\def\showchanges{0} 
\newcommand{\lund}{Department of Physics, Lund University, P. O. Box 118, SE-22100 Lund, Sweden}

\newcommand{\gu}{Department of Physics, University of Gothenburg, Origov\"agen 6B, SE-41296 Gothenburg, Sweden}
 \newcommand{\uam}{Departamento de Qu\'imica, M\'odulo 13, Universidad Aut\'onoma de Madrid, 28049 Madrid, Spain}
\newcommand{\saclay}{LIDYL, CEA, CNRS, Universit\'e Paris-Saclay, CEA Saclay, 91191 Gif-Sur-Yvette, France}
\newcommand{\IMDEA}{Instituto Madrile\~no de Estudios Avanzados en Nanociencia (IMDEA-Nanociencia), Cantoblanco, 28049 Madrid, Spain}
\newcommand{\IFIMAC}{Condensed Matter Physics Center (IFIMAC), Universidad Autónoma de Madrid, 28049 Madrid, Spain}
\newcommand{\UCF}{Department of Physics \& CREOL, University of Central Florida, FL32816, USA}
\newcommand{\MBI}{Max-Born Institute for Nonlinear Optics and Short Pulse Spectroscopy, Max-Born-Stra{\ss}e 2A, D-12489 Berlin, Germany}

\documentclass[10pt]{iopart}

\usepackage{epsfig,amsbsy,bm}
\usepackage{etoolbox}
\usepackage{color}
\usepackage{placeins}
\usepackage{mathrsfs}
\usepackage{iopams}
\usepackage[colorlinks,linkcolor=red,citecolor=blue]{hyperref}
\usepackage{cancel}
\usepackage{graphicx}
\usepackage{graphicx,color}
\usepackage{braket}
\usepackage{upgreek}
\usepackage{dcolumn}
\usepackage[nice]{nicefrac}
\usepackage[tight]{units}
\usepackage{lipsum}
\usepackage{ulem}
\usepackage[dvipsnames]{xcolor}
\usepackage{pgfplots}
\usepackage[utf8]{inputenc}
\usetikzlibrary{patterns}
\usetikzlibrary{arrows.meta}
\usetikzlibrary{fadings}
\usetikzlibrary{decorations.markings}
\usetikzlibrary{shapes,arrows}
\tikzstyle{block} = [draw, fill=white, rectangle,
    minimum height=3em, minimum width=6em]
\tikzstyle{sum} = [draw, fill=white, circle, node distance=1cm]
\tikzstyle{input} = [coordinate]
\tikzstyle{output} = [coordinate]
\tikzstyle{pinstyle} = [pin edge={to-,thin,black}]
\tikzset{>={Latex[width=2mm,length=2mm]}}
\pgfplotsset{compat=newest}
\pgfplotsset{plot coordinates/math parser=false}
\usepgfplotslibrary{external}
\tikzset{external/force remake}
\definecolor{changes}{rgb}{0.0,0.0,0.0}%
\ifnum\showchanges=1 %
\definecolor{changes}{rgb}{0.0,0.0,1.0}%
\fi

\begin{document}
\title{Time-frequency representation of autoionization dynamics in helium}

\author{D. Busto$^1$, L. Barreau$^2$, M. Isinger$^1$, M. Turconi$^2$, C. Alexandridi$^2$, A. Harth$^1$, S. Zhong$^1$, R. J. Squibb$^3$, D. Kroon$^1$, S. Plogmaker$^1$, M. Miranda$^1$, \'A. Jim\'enez-Gal\'an$^4$, L. Argenti$^5$, C. L. Arnold$^1$, R. Feifel$^3$, F. Mart\'in$^{6,7,8}$, M. Gisselbrecht$^1$, A. L'Huillier$^1$, P. Sali\`eres$^2$}

\address{$^1$\lund}
\address{$^2$\saclay}
\address{$^3$\gu}
\address{$^4$\MBI}
\address{$^5$\UCF}
\address{$^6$\uam}
\address{$^7$\IMDEA}
\address{$^8$\IFIMAC}
\ead{david.busto@fysik.lth.se}
\vspace{10pt}

\begin{abstract}
Autoionization, which results from the interference between direct photoionization and photoexcitation to a discrete state decaying to the continuum by configuration interaction, is a well known example of the important role of electron correlation in light-matter interaction.
Information on this process can be obtained by studying the spectral, or equivalently, temporal complex amplitude of the ionized electron wavepacket. Using an energy-resolved interferometric technique, we measure the spectral amplitude and phase of autoionized wavepackets emitted via the sp2$^+$ and sp3$^+$ resonances in helium. These measurements allow us to reconstruct the corresponding temporal profiles by Fourier transform. In addition, applying various time-frequency representations, we observe the build up of the wavepackets in the continuum, monitor the instantaneous frequencies emitted at any time and disentangle the dynamics of the direct and resonant ionization channels.
\end{abstract}

\maketitle
\ioptwocol

\section{Introduction}

Upon the absorption of a sufficiently high energy photon, an electron in a bound system can be ionized. The escaping electron may interact with the remaining electrons leading to various multi-electronic processes such as shake-up, double ionization via shake-off or knock-out, or Auger decay (see \cite{SchmidtRPP92} for an historical review). Another ionization mechanism induced by electron-electron interaction is autoionization, which results from excitation to a quasi-bound state which decays to the continuum. Autoionization, theoretically described in a seminal article by U. Fano \cite{FanoPR1961}, is a quantum interference effect between the direct path to the continuum and the resonant path through the quasi-bound state. The interference leads to the famous asymmetric Fano profile, characterized by a transition amplitude given by
\begin{equation}
R(\epsilon,q)=\frac{q+\epsilon}{\epsilon+i},
\label{eq:resonant_factor}
\end{equation}
where \textcolor{changes}{$i$ is the imaginary unit,} $q$ is the asymmetry parameter, proportional to the ratio between the direct and resonant transition amplitudes and $\epsilon$ is the reduced energy $\epsilon = 2(E-E_\mathrm{\Phi})/\Gamma$. Here $E$ is the continuum energy, $E_\mathrm{\Phi}$ the energy of the quasi-bound state and $\Gamma$ its spectral width.

A textbook example of simple systems exhibiting electron correlations is the $^1$P$^o$ series of doubly excited states in He converging to the $N$ = 2 state of He$^+$. Their observation by Madden and Codling \cite{MaddenCodlingPRL1963} in 1963 indicated the breakdown of the independent electron picture, leading to strong theoretical activity to understand correlated two-electron dynamics (e.g. \cite{TannerRevModPhys2000} and references therein). Experimentally, high resolution spectroscopic studies at synchrotron facilities have led to the determination of accurate spectroscopic parameters of a few Rydberg series \cite{MorganPRA1984,KossmannJPhysB1988,DomkePRA1996}. More recently, with the increasing quality of experimental techniques, intriguing aspects often overlooked have been discussed such as the competition between autoionization and fluorescence decays \cite{RubenssonPRL1999}, the role of relativistic effects \cite{PenentPRL2001}, and possible mechanism for double excitation \cite{CzashPRL2005}.

The development of attosecond science has brought new insight into multi-electronic processes \cite{OssianderNatPhys2016,IsingerScience2017, ManssonNatPhys2014,DrescherNature2002}, and in particular opened up the possibility to measure real time dynamics of autoionizing states with pump-probe methods. Using the attosecond streaking technique \cite{WickenhauserPRL2005,ZhaoPRA2005}, the lifetime of the lowest doubly excited state in helium was determined \cite{GilbertsonPRL2010}. An autoionizing decay of 8 fs was measured in Ar \cite{WangPRL2010} using attosecond transient absorption.  These two results were found in very good agreement with spectroscopic data. However, the lifetime is not sufficient to describe the entire autoionization dynamics, in particular the interferences between the direct and resonant ionization paths that are responsible for the asymmetric Fano lineshape. These interferences were not observed in the above experiments, probably because they occur shortly after the initial excitation, and were smoothed out by the ~8-fs infrared probe beam.

A spectral approach provides an alternative to direct measurements in the time domain. The information on the autoionization dynamics, encoded in the complex spectral transition amplitude [Eq. \ref{eq:resonant_factor}], requires the measurement of its amplitude $|R(\epsilon,q)|$ and phase:
\begin{equation}
\arg \left[ R(\epsilon,q) \right] = \arctan \epsilon - \pi \Theta(\epsilon + q) + \frac{\pi}{2}
\label{eq:argRe}
\end{equation}
where $\Theta$ is the Heaviside function.
Spectral phase measurements can be performed by combining a comb of high harmonics (a train of attosecond pulses in the time domain) and a weak ($\approx$10$^{11}$ W/cm$^2$) IR probe, with the so-called RABBIT technique (Reconstruction of Attosecond Beating by Interference of Two-photon Transitions) \cite{PaulScience2001,MullerApplPhysB2002}. This method (together with its generalized FROG-CRAB version \cite{MairessePRA2005}) has allowed observing the signature of phase distortions induced by autoionizing resonances  \cite{HaesslerPRA2009,HaesslerNJP2013,SabbarPRL2015}. Recently, the spectral phase variation induced by an autoionizing resonance in argon was measured by scanning the harmonic frequency across the resonance and recording RABBIT traces for each frequency \cite{KoturNatCom2016}. Using a spectrally-resolved technique, which we refer to as Rainbow RABBIT, Gruson and coworkers fully characterized the electron wave packet emitted through the first autoionizing state in helium and could thus reconstruct the build-up of the resonance profile in the time domain \cite{GrusonScience2016}. Similar build-up was obtained by transient absorption spectroscopy \cite{KaldunScience2016} using an intense ($10^{13}$ W/cm$^2$) probe pulse providing a fast gate. Control of the Fano profile was also demonstrated by varying the intensity of this probe pulse \cite{OttScience2013,OttNature2014}.

Here, we  characterize electronic wave packets emitted in He in the vicinity of two doubly excited states, denoted sp2$^+$ and sp3$^+$ (based on Cooper's classification\cite{CooperPRL63}), whose energy, asymmetry parameter, linewidth and lifetime $\tau = \hbar /\Gamma$ are summarized in table \ref{tab:FanoParameters}.
We use the Rainbow RABBIT method with a tunable titanium sapphire laser system to study both the spectral amplitude and phase of the resonant electron wave packets (EWPs).  
\textcolor{changes}{This paper aims at pushing forward the analysis of autoionization dynamics in helium that was presented in \cite{GrusonScience2016}. The experiments are performed in different experimental conditions, over a broader energy range (including, e.g. sp3+). The influence of the different experimental parameters such as the spectrometer resolution and the spectral width of the IR and XUV pulses is discussed in details. Finally, we investigate different time frequency representations and, in the case of the sp2$^+$ resonance, we fully characterize the resonant EWP using Short Time Fourier Transforms (STFT) and Wigner time-frequency representations. This, together with theoretical calculations, allows us to resolve the ionization dynamics, and in particular, to disentangle the contributions of the two ionization paths.} 

\begin{table}
\begin{tabular}{|c|c|c|c|c|}
\hline
Resonance & $E_\mathrm{\Phi}$ (eV) & $q$ & $\Gamma$ (meV) & $\tau$ (fs) \\
\hline
He sp2$^+$ & 60.15 & -2.77 & 36 & 17 \\
\hline
He sp3$^+$ & 63.66 & -2.58 & 8 & 82\\
\hline
\end{tabular}
\caption{Energy $E_\mathrm{\Phi}$, asymmetry parameter $q$, spectral width $\Gamma$ and lifetime $\tau$ of the two autoionizing states of this study. Spectroscopic data from \cite{DomkePRA1996}.}
\label{tab:FanoParameters}
\end{table}

This paper is structured as follows. In section \ref{sec:methods}, we present the experimental setup and methods used. Section \ref{sec:limitations} discusses the limitations to our measurements. In section \ref{sec:results}, the results are presented and compared to theoretical calculations. Finally, section \ref{sec:time_frequency} is devoted to the representation of the autoionization dynamics in the time-frequency domain.

\section{Methods}
\label{sec:methods}
\subsection{Experimental setup}
\label{subsec:experimental_setup}
\begin{figure}[t]
  \includegraphics{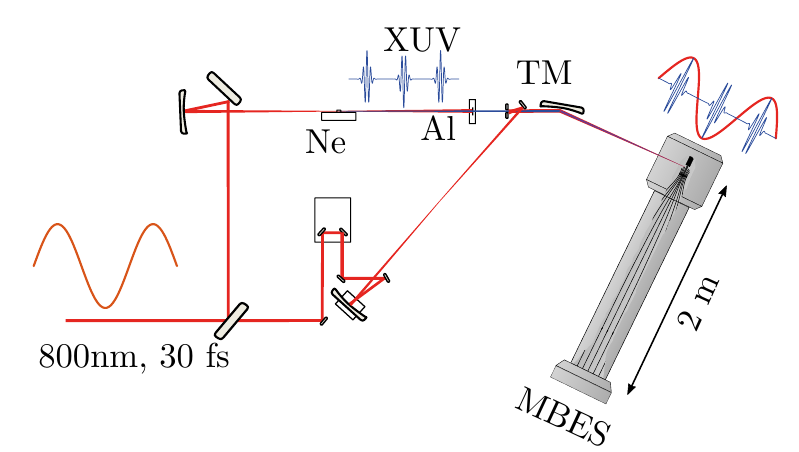}
  \caption{\textcolor{changes}{Schematic drawing of the experimental setup. The driving laser is split in the interferometer, where 70 \% is sent to the top arm in the drawing. The beam is focused into a gas cell filled with neon and filtered out with the Al foil. The emerging XUV pulses are sent through the center of a drilled mirror and focused by the toroidal mirror (TM) in the interaction region of the MBES. The other 30 \% is sent through a variable delay stage and focused in vacuum, to match the wavefront of the XUV, before it is reflected by the outer annular part of the drilled mirror and then focused by the toroidal mirror.}}
  \label{fig:setup}
\end{figure}

The experiments were performed with a 1 kHz titanium sapphire laser producing pulses centered at 800 nm with a spectral width of 85 nm and a pulse duration around 22 fs. A dazzler was used to shape the pulse spectrum, allowing for the tuning of the central frequency from 790 nm to 810 nm with a reduced bandwidth of 65 nm.  The pulses\textcolor{changes}{, now 30 fs long,} were sent to a spatially- and temporally-stabilized Mach-Zehnder interferometer \cite{KroonOL2014} where the pulses were split in two arms as shown in figure \ref{fig:setup}. In the first arm they were focused with an on-axis spherical mirror ($f=50$ cm focal length) in a 10-mm long gas cell filled with neon to generate high-order harmonics reaching energies in the extreme ultra-violet (XUV) up to 110 eV. A 200-nm thick aluminium foil was placed after the generating medium to filter out the remaining IR as well as the harmonics above 70 eV. The tunability of the laser source allowed us to choose particular harmonics to excite different autoionizing states, here the  sp2$^+$ or the sp3$^+$ Fano resonances in helium located 60.15 and 63.66 eV from the ground state, respectively (table \ref{tab:FanoParameters}). In the second arm, a $\lambda/2$ wave plate and a broad-band polarizer were used to control the intensity of the probe beam which was delayed from the generated XUV pulse train with a piezoelectric stage. Both arms were then recombined \textcolor{changes}{with a drilled flat mirror} and focused with a toroidal mirror ($f=30$ cm) in a helium gas jet where they were spatially and temporally overlapped. The photoelectron spectrum resulting from the interaction of the two pulses with the helium atoms was measured \textcolor{changes}{over a sequence of delays} with a 2-meter-long Magnetic Bottle Electron Spectrometer (MBES)\textcolor{changes}{. The spectrometer} combines $4\pi$ sr collection efficiency with sub-100 meV resolution for electrons with low kinetic energy (below 10 eV). A retarding potential was applied to shift the photoelectron spectrum in this energy region.\textcolor{changes}{Our measurements are in the spectral domain, and the temporal information is obtained by Fourier transformation. Our temporal resolution is thus solely given by the inverse of the spectral width of the EWPs. We estimate our temporal resolution to be around 2 fs.}

\subsection{RABBIT method}
Figure \ref{fig:rabbit}(b) shows a section of the delay-integrated photoelectron spectrum which is composed of a set of intense peaks (H39, H41) spaced by $2\hbar\omega_0$, where $\omega_0$ is the central frequency of the IR field, resulting from the photoionization of the atoms by the harmonic comb. In the presence of the weak IR field  \textcolor{changes}{($\sim 10^{11}$W/cm$^2$)}, two-photon transitions (XUV$\pm$IR) can occur, as illustrated in figure \ref{fig:rabbit}(a), leading to the appearance of sidebands in between the harmonic peaks (SB38, SB40, SB42) \cite{VeniardPRA1996,MairesseScience2003}. The energy scale is the total photon energy absorbed in the process (this will be used throughout the article). As shown in figure \ref{fig:rabbit}(c), when the delay $\tau$ of the IR field relative to the XUV is varied, the intensity of the sidebands oscillates at twice the fundamental frequency according to:

\begin{eqnarray}
 \label{eq:SB}
I_\mathrm{n+1}(E_\mathrm{f},\tau)=&|A_\mathrm{n+2-1}(E_\mathrm{f})|^2+|A_\mathrm{n+1}(E_\mathrm{f})|^2\\
&+2|A_\mathrm{n+2-1}(E_\mathrm{f})||A_\mathrm{n+1}(E_\mathrm{f})| \nonumber \\
&\times \cos[2\omega_0\tau-\Delta\phi_\mathrm{XUV}(E_\mathrm{f})-\Delta\varphi_\mathrm{A}(E_\mathrm{f})] \nonumber,
\end{eqnarray}
where $E_\mathrm{f}$ is the total energy absorbed, $A_\mathrm{n+1}$ ($A_\mathrm{n+2-1}$) is the complex two-photon transition amplitude corresponding to the absorption of harmonic H$_\mathrm{n}$ (H$_\mathrm{n+2}$) and the absorption (emission) of one IR photon, respectively, $\Delta\phi_\mathrm{XUV}$ is the phase difference between two consecutive harmonics, and $\Delta\varphi_\mathrm{A}$ is the phase difference between the two-photon transition dipole matrix elements. \textcolor{changes}{In the experiment, the delay was varied over 12 fs around the zero delay (overlap).}

The usual implementation of the RABBIT technique consists in integrating $I_\mathrm{n+1}(E_\mathrm{f},\tau)$ over energy inside each sideband $n+1$ and extract the phase of the $2\omega_0$ oscillations, giving direct access to $\Delta\phi_\mathrm{XUV}+\Delta\varphi_\mathrm{A}$ \cite{PaulScience2001}.
Due to the generation mechanism of the high-order harmonics, the XUV pulse train carries an intrinsic quadratic phase, the attochirp, which leads to the approximately linear increasing phase difference $\Delta\phi_\mathrm{XUV}=\phi_\mathrm{n+2}-\phi_\mathrm{n}$ between consecutive harmonics \cite{MairesseScience2003}. The aluminum foil used to filter out the IR after the gas cell partly compensates for this effect \cite{LopezMartensPRL2005}. The phase $\Delta\varphi_\mathrm{A}$ arises from the two-photon ionization process. In the case of non-resonant ionization, this phase is smoothly varying \cite{KlunderPRL2011}. In contrast, when one of the paths goes through a bound or quasibound intermediate state {\cite{SwobodaPRL2010,CaillatPRL2011}, $\Delta\varphi_\mathrm{A}$ strongly varies with the detuning from the resonance.
For instance, when H39 is resonant with sp2$^+$ (see figure \ref{fig:rabbit}), $\Delta\varphi_\mathrm{A}$   for SB38 and SB40 is affected, while $\Delta\phi_\mathrm{XUV}$ remains the same. By scanning H39 across the resonance and recording the corresponding RABBIT traces, the spectral variation of $\Delta\varphi_\mathrm{A}$ can be recovered. Note, however, that the integration over the sidebands results in a mean value of $\Delta\varphi_\mathrm{A}$ over the harmonic bandwidth. In our conditions, the spectral bandwidth of the generated harmonics, equal to 180 meV, is much larger than the sp2$^+$ resonance width (table \ref{tab:FanoParameters}), which results in a strong smoothing of the spectral evolution of $\Delta\varphi_\mathrm{A}$ (see supplementary material of \cite{GrusonScience2016}).

  \begin{figure}[t]
\begin{center}
    \includegraphics{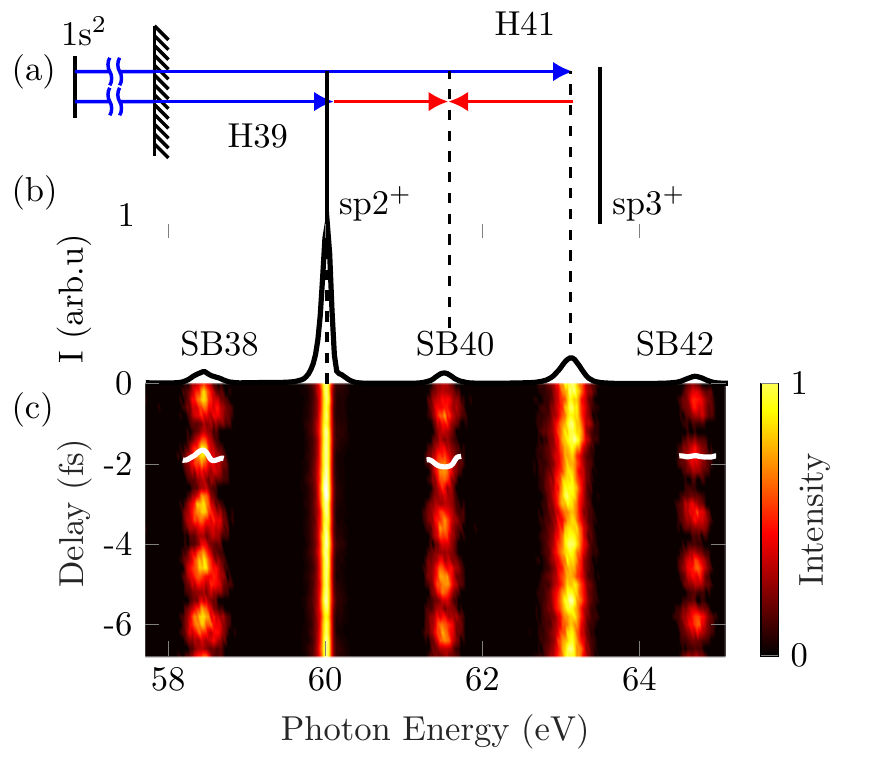}
\end{center}
 \caption{(a) Energy diagram and transitions when harmonic 39 is resonant with the sp2$^+$ state. The blue arrows correspond to the ionization from the ground state (1s$^2$) by the different harmonics, while the red arrows show the IR transitions used to couple the consecutive harmonics and giving rise to the sidebands. (b) Normalized intensity of the delay-integrated photoelectron spectrum showing the signature of the sp2$^+$ resonance in harmonic 39. (c) Photoelectron spectra taken as a function of the relative delay between the XUV pulse-train and the IR pulse. The intensity of the sidebands has been multiplied by 7 and the intensity of H41 by 5 for the sake of visibility. The intensity of the delay scan has been normalized to the maximum intensity of the whole scan. The Fano phase is imprinted in the phase of the neighboring sidebands 38 and 40, as shown by the white lines while the nonresonant sideband (SB42) has a flat phase. In this case, harmonic 41 is not resonant with the sp3$^+$ state, leading to sideband 42 being unaffected.}
 \label{fig:rabbit}
 \end{figure}
\subsection{Rainbow RABBIT}

The high spectral resolution of our spectrometer allows us to use the Rainbow RABBIT technique \cite{GrusonScience2016} to directly retrieve the phase variation across the resonance, i.e. $\Delta\varphi_\mathrm{A}(E_\mathrm{f})$, by analyzing the $2\omega_0$ oscillations at each energy $E_\mathrm{f}$ inside the sidebands (see white lines in figure \ref{fig:rabbit}(c)). A single RABBIT trace in resonant conditions may then give access to the full phase information around the resonance.

We compared two different techniques to extract the phase of the sidebands. The first one relies on the fitting of the oscillations of the sidebands based on equation \ref{eq:SB} for each energy. The second one is based on the extraction of the phase of the Fourier-transformed oscillations. Both methods give almost identical results. All the phase measurements presented in this article were obtained using the fitting technique.

To extract the amplitude of the resonant two-photon EWP, we Fourier transform the sideband intensity $I_\mathrm{n+1}(E_\mathrm{f},\tau)$ and extract the signal oscillating at frequency $2\omega_0$ [$\Delta I_\mathrm{n+1}(E_\mathrm{f})$]. This allows us to eliminate the first two terms in equation (\ref{eq:SB}) which are delay-independent, and contribute to noise. $\Delta I_\mathrm{n+1}(E_\mathrm{f})$ is the product of the amplitude of the resonant transition with the amplitude of the non resonant transition. For example, for sideband 38, $\Delta I_{38}(E_\mathrm{f})=2|A_{37+1}(E_\mathrm{f})||A_{39-1}(E_\mathrm{f})|$ where $A_{39-1}$ is the amplitude of the resonant two-photon transition and $A_{37+1}$ is the amplitude of the non resonant transition which is used as reference. To isolate only the resonant amplitude, i.e. $|A_{39-1}(E_\mathrm{f})|$, the same procedure is performed with a non resonant sideband, SB44 (not visible in figure \ref{fig:rabbit}), where both paths are non resonant. Assuming that all the non resonant amplitudes are similar, $A_{37+1}\approx A_{43+1}\approx A_{45-1}$, the amplitude of the resonant EWP can be extracted according to \cite{GrusonScience2016}:
\begin{equation}
|A_{39-1}|=\frac{\Delta I_{38}}{\sqrt{2\Delta I_{44}}}.
\label{eq:calibration}
\end{equation}
\textcolor{changes}{This approximation is justified by the fact that these harmonics are in the plateau region and that the transmission of the aluminum filter is approximately constant in this energy range.}

\subsection{Theoretical description}
\label{subsec:theory}

We use an analytical model introduced in \cite{GalanPRA2016}, whose validity has been checked against fully correlated \textit{ab-initio} calculations \cite{GalanPRL2014}. The interaction between the IR and XUV pulses and the atom is treated within the framework of second-order time-dependent perturbation theory. The spectral amplitudes of the light fields are given by
\begin{eqnarray}
\mathcal{E}_\mathrm{IR}(\omega,\tau)&= \mathcal{E}^0_\mathrm{IR} e^{i\omega_0\tau} \exp\left[-\frac{(\omega-\omega_0)^2}{2\sigma_\mathrm{IR}^2}\right] \nonumber\\
\mathcal{E}_\mathrm{XUV}(\Omega)&= \mathcal{E}^0_\mathrm{XUV} \exp\left[-\frac{(\Omega-\Omega_0)^2}{2\sigma_\mathrm{XUV}^2}\right],
\end{eqnarray}
where $\sigma_\mathrm{IR}$, $\sigma_\mathrm{XUV}$ are the bandwidths of the IR and XUV fields respectively. $\omega$, $\Omega$ denote the IR and XUV photon frequencies,  while $\omega_0,\Omega_0$ are the respective central frequencies. $\mathcal{E}^0_\mathrm{IR}$ and $\mathcal{E}^0_\mathrm{XUV}$ are constant amplitudes.
The two-photon transition amplitude which takes into account the extended bandwidths of both IR and XUV pulses, can be approximated, in the case of IR photon absorption, as
\begin{equation}
A_\mathrm{n+1}(E_\mathrm{f})=-i\hbar \!\!\int_0^\infty \!\!\!\!\!\!\!  d\Omega\  \mathcal{E}_\mathrm{IR}(\Omega_\mathrm{fg}-\Omega,\tau)\mathcal{E}_\mathrm{XUV}(\Omega)M_\mathrm{fg}(\Omega),
\label{eq:2photon_amplitude}
\end{equation}
where $\Omega_\mathrm{fg}=(E_\mathrm{f}-E_\mathrm{g})/\hbar$. The index $n+1$ indicates that we consider absorption of the $n$th harmonic plus one IR photon (see equation \ref{eq:SB}). The two-photon transition matrix element $M_\mathrm{fg}$ can be written as:
 \begin{equation}
M_\mathrm{fg}(\Omega) \propto \sum\kern-1.4em\int dE
\frac{\braket{\psi_{E_\mathrm{f}}|T|\Psi_E}\braket{\Psi_E|T|g}}{E_\mathrm{g}+\hbar\Omega-E+i0^+},
\label{eq:Mn}
 \end{equation}
where $\ket{g}$ is the ground state 1s$^2$, $\ket{\Psi_E}$ is the intermediate state and $\ket{\psi_{E_\mathrm{f}}}$ is the final state. These states are respectively at the energies $E_\mathrm{g}$, $E$ and $E_\mathrm{f}$ and are coupled by the dipole transition operator $T$. The calculation of the two-photon transition matrix element requires to sum over all the intermediate discrete and continuum states.

Our theoretical derivation follows the well-known formalism developed by U. Fano \cite{FanoPR1961} to account for the interaction between the continuum channels and the quasi-bound states and generalizes it to include the influence of a weak IR field, in the perturbative limit. Fano's theoretical approach consists in calculating the eigenstates of  $H=H_0+V$, where $V$ couples the bound state $\ket{\varphi}$ and the nonresonant continuum states $\ket{\psi_\mathrm{E}}$ of the unperturbed Hamiltonian $H_0$. Following Fano's formalism, one can easily relate the eigenstates of  $H$, $\ket{\Psi_\mathrm{E}}$ with those of $H_0$, $\ket{\varphi}$  and $\ket{\psi_\mathrm{E}}$ \cite{FanoPR1961}.
The asymmetry parameter $q$ introduced in equation (1) is equal to
\begin{equation}
q =\frac{\langle \Phi|T| g\rangle }{\pi V^{*}_\mathrm{E} \langle \psi_\mathrm{E}|T| g\rangle},
\end{equation}
where $V_\mathrm{E}= \langle \psi_\mathrm{E}|V|\varphi\rangle $ and $\ket{\Phi}$ is the bound state $\ket{\varphi}$ modified by the configuration interaction \cite{FanoPR1961}. The autoionizing state bandwidth $\Gamma$ is equal to $2\pi|V_\mathrm{E}|^2$.

As shown in \cite{GalanPRA2016}, the effect of an autoionizing resonance in the intermediate state of a two-photon absorption process can be taken into account by multiplying the nonresonant two-photon matrix element $M_\mathrm{fg}$ calculated using unperturbed wavefunctions $\ket{\psi_\mathrm{E}}$ by $R(\epsilon,q_\mathrm{eff})$, which includes the effect of the resonance. The effective parameter $q_\mathrm{eff}$ is complex and equal to $q(1-\gamma) +i\gamma$, where $\gamma$ is a real parameter that depends on the relative strength of the direct transition from the intermediate bound state to the final continuum versus the nonradiative coupling of the bound state to the intermediate continuum state followed by the dipole coupling to the final continuum. Both $\epsilon$ and $q$ are calculated at the energy $E_\mathrm{g}+\hbar \Omega$.
For two-electron transitions, as is the case for transitions from spn$^+$ to 1s$\epsilon$s or 1s$\epsilon$d, $\gamma$ is usually small \cite{GalanPRA2016}. In the present work, we used $\gamma_\mathrm{sp2^+}=-0.025$, $\gamma_\mathrm{sp3^+,\epsilon s}=-0.117$, $\gamma_\mathrm{sp3^+,\epsilon d}=-0.390$.
\section{Experimental limitations}
\label{sec:limitations}

In this section, we discuss limitations which are inherent to the measurement process.

\subsection{Spectrometer resolution}
\label{subsec:spectometer}

For each delay, the measured time of flight spectrum is a convolution of the photoelectron spectrum with the electron spectrometer response function (RF), which limits our spectral resolution. In this work, we implement a deconvolution algorithm to retrieve the real photoelectron spectrum. We assume an energy independent spectrometer response within the range of study and apply  a two-dimensional iterative blind deconvolution algorithm based on the maximum likelihood Lucy-Richardson (LR) method \cite{RichardsonJOSA1972,LucyAstronJ1974}. Given a measured (convoluted) spectrum $S$, a blind algorithm attempts to find the real photoelectron spectrum $s$ and the RF $f$ such that $S=f\otimes s +n$, where $n$ is noise contamination and $\otimes$ is the convolution operator. The algorithm starts from an initial estimate of $s$ and $f$. For each cycle, multiple Lucy-Richardson iterations are performed,
\begin{eqnarray}
f_{k+1}&=\frac{1}{\sum s_{k}}f_{k} s_{k}\star \left(\frac{S}{f_{k}\otimes s_{k}}\right) \nonumber \\
s_{k+1}&=\frac{1}{\sum f_{k+1}}s_{k} f_{k+1}\star \left(\frac{S}{f_{k+1}\otimes s_{k}}\right)
 \end{eqnarray}
where $\star$ is the correlation operator and the symbol $\sum$ denotes the spectral integral. We impose the constraint that the RF should be Gaussian. For each cycle, the likelihood that the retrieved quantities reproduce the measured spectrum by a convolution increases. Further information about the algorithm can be found in \cite{BiggsSPIE1998}.
The RFs retrieved for the different spectra were similar, with a spectral width of $89\pm$5 meV at full width half maximum, close to the estimated experimental resolution. This shows the stability of the deconvolution algorithm.

 \begin{figure}[t]
 \begin{center}
 \includegraphics{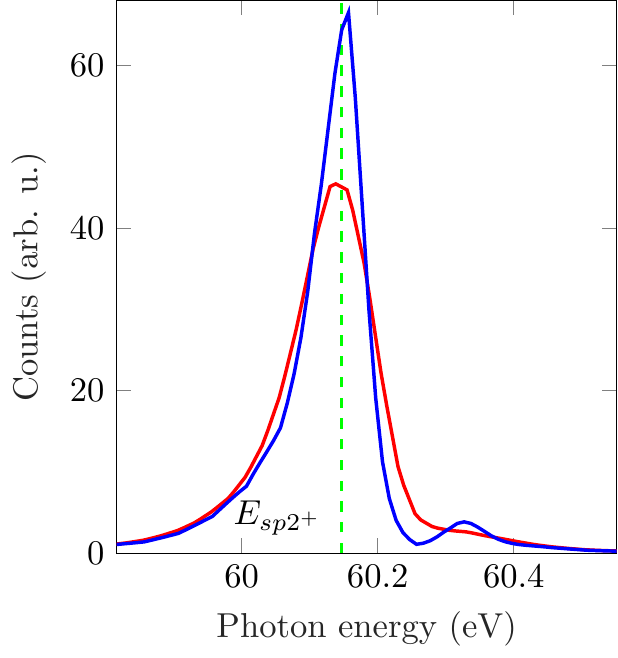}
 \end{center}
 \caption{Deconvolution of the photoelectron spectrum. Profile of the resonant harmonic (H39) before (red) and after (blue) deconvolution. The green dashed line indicates the position of the resonance.}
 \label{fig:deconv_H39}
 \end{figure}

Figure \ref{fig:deconv_H39} shows the photoelectron spectrum obtained by absorption of the 39th harmonic, close to the sp2$^+$ resonance, before and after deconvolution (respectively red and blue curves). The deconvolution reduces the width of the harmonic and enhances the characteristic asymmetry of the Fano profile.
The experimental deconvolved spectrum shows a minimum after the resonance and a second peak at higher energies which corresponds to the nonresonant spectrum of the harmonic \cite{GrusonScience2016}.

 \begin{figure}[t]
 \includegraphics{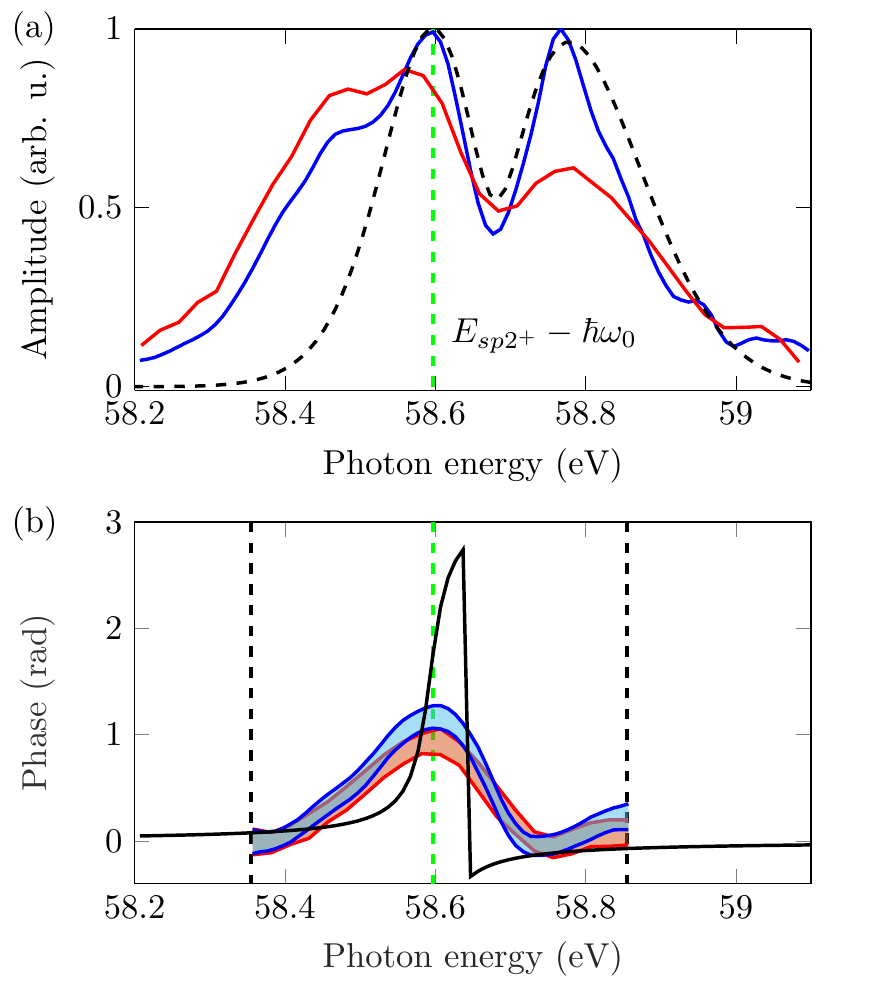}
     \caption{Amplitude and phase of the two-photon electron wave packet ($A_{39-1}$) emitted when H39 is resonant with sp2$^+$. (a) Normalized amplitude before (red line) and after (blue line) deconvolution. The latter is a good agreement with the simulations (black dashed line). The green dashed line indicates the position of $E_\mathrm{sp2^+}-\hbar\omega_0$. (b) Spectral phase before (red) and after (blue) deconvolution. We set a threshold of $30\%$ of the maximum intensity of the sideband as a limit below which we consider that the intensity to noise ratio is too low to reliably extract the phase. The black dashed lines correspond to the limits of the energy region in which the intensity of the sideband is above the threshold. The shaded area represents the standard deviation given by the fit. The black solid line corresponds to the phase of the resonant one-photon transition amplitude (see equation \ref{eq:argRe}), shifted down by one laser photon energy. }
      \label{fig:deconv}
  \end{figure}

We applied the deconvolution algorithm to the full RABBIT trace and extracted the amplitude and phase of the sidebands from the new spectrogram. In figure \ref{fig:deconv}, we compare the amplitude (a) $|A_{39-1}|$ calculated using equation \ref{eq:calibration}, and phase (b) of the two-photon transition before (red) and after (blue) deconvolution, in conditions such that H39 is resonant with the sp2$^+$ state.
The deconvolution gives sharper features in figure \ref{fig:deconv}(a), which agree well with theoretical results indicated by the dashed black line (see Section \ref{subsec:theory} and \ref{subsec:spectral_domain}). We could not identify the spectral feature observed around 58.4 eV. For the phase variation, the deconvolution leads      to a slightly sharper phase evolution around the resonance.
However, the latter is still quite different from the resonant one-photon dipole phase shown as black line in figure \ref{fig:deconv}(b). According to equation \ref{eq:argRe}, this phase displays a smooth $\pi$  variation characteristic of the $\arctan$ function around $\epsilon = 0$, followed by a sudden $\pi$ phase jump at $\epsilon = -q$. The reason for the difference is thus not the convolution with the RF of the spectrometer, as was the case in \cite{GrusonScience2016}, but finite pulse effects, as discussed below.

\subsection{Finite pulse effects}
\label{subsec:finitepulses}

It is often considered that in the RABBIT scheme, the IR field makes a perfect replica of the wave packet excited by the harmonic and that the amplitude and phase measured in the sideband correspond to that of the one-photon wave packet. While in the case of long pulses with narrow spectra, the correspondence between one- and two-photon wave packets is justified, this approximation breaks down when the bandwidths of the IR and XUV pulses become large. Indeed, in the presence of broad pulses, multiple combinations of frequencies can lead to the same final state,
thus giving rise to a coherent mixing of the different frequencies of the one-photon wave packet. In \cite{GalanPRA2016}, Jim\'enez-Gal\'an \textit{et al.} showed that this  mixing of frequency components can induce a variety of effects, referred to as finite pulse effects, ranging from a smoothing of the amplitude and phase of the two-photon EWP to a modification of the oscillation frequency of the sidebands. For nonresonant transitions, such that $M_\mathrm{fg}$ does not depend on the frequency over the pulse bandwidth, equation \ref{eq:2photon_amplitude} is a convolution of the one-photon wave packet with the IR pulse. On the contrary, in the case of ionization through an autoionizing state, due to the strong frequency dependence of $M_\mathrm{fg}(\Omega)$, the two-photon wave packet cannot be approximated as the convolution of the one-photon wave packet with the IR pulse. This leads to a smoothing of both amplitude and phase which cannot be corrected by the deconvolution algorithm.
 In our experimental conditions, the large bandwidth of the IR pulse (125 meV) prevents us from approximating the measured amplitude and phase to those of the one-photon wave packet as evidenced in figure \ref{fig:deconv}(b).

\subsection{Harmonic blueshift}
\label{subsec:blueshift}

As described in section \ref{subsec:experimental_setup}, the high-order harmonics are produced by focusing (part of) the laser beam in a gas cell. The laser intensity is high enough so that the front of the pulse can partially ionize the medium, thus creating a low density plasma in which the pulse propagates. The interaction of the IR pulse with the plasma leads to a blue shift of the laser central frequency that results in harmonics separated in frequency by $2(\omega_0+\delta_\omega)$ \cite{WahlstromPRA1993}. Since the probe IR pulse does not propagate
through the gas cell and is thus not blue-shifted, the contributions from the lower and higher harmonics to the sideband do not perfectly overlap in frequency.
In the absence of blue shift, the quadratic phase variation \textit{inside} the harmonic lines (due to the harmonic chirp, not to be confused with the atto-chirp \cite{SalieresPRL1995,VarjuJMO2005}) does not influence our measurement. Indeed, the variations of $\phi_\mathrm{n+2}$ and $\phi_\mathrm{n}$ are similar over the pulse bandwidth so that $\Delta\phi_{XUV}$ only contributes to a constant phase in equation \ref{eq:SB}. In the presence of a blue shift,
$\Delta\phi_{XUV}$ varies linearly with frequency, with a coefficient equal to $-8\delta_\omega \phi''_\mathrm{n}$, where $\phi''_\mathrm{n}$ is the harmonic group delay dispersion.
In the experimental results, a linear phase variation was indeed observed for the nonresonant sidebands. This linear phase is removed in all the results presented below.

\section{Results}
\label{sec:results}

\subsection{Spectral domain}
\label{subsec:spectral_domain}

\begin{figure}[t]
 \centering
   \includegraphics{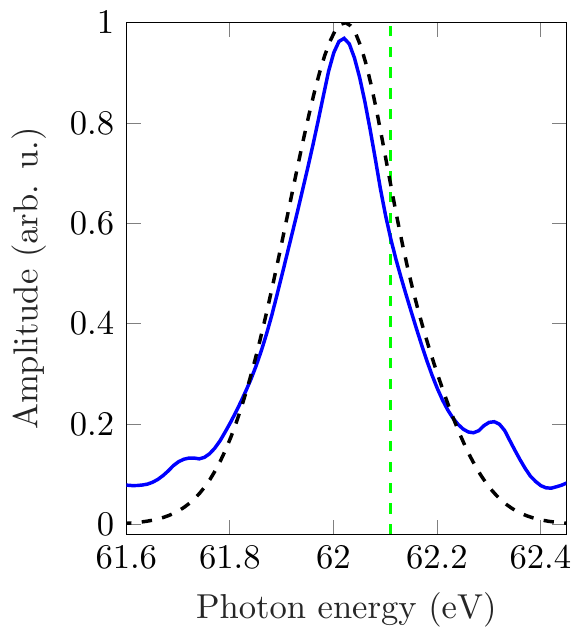}
     \caption{\textcolor{changes}{Amplitude of the two-photon electron wave packet ($A_{41-1}$) emitted when H41 is resonant with the sp3$^+$ state. The blue curve corresponds to the measured amplitude and agrees well with the simulation (black dashed curve). The green dashed line indicates the position of $E_{\mathrm{sp3^+}}-\hbar\omega_0$.}}
      \label{fig:sp3}
  \end{figure}

 \begin{figure*}[t]
   \includegraphics{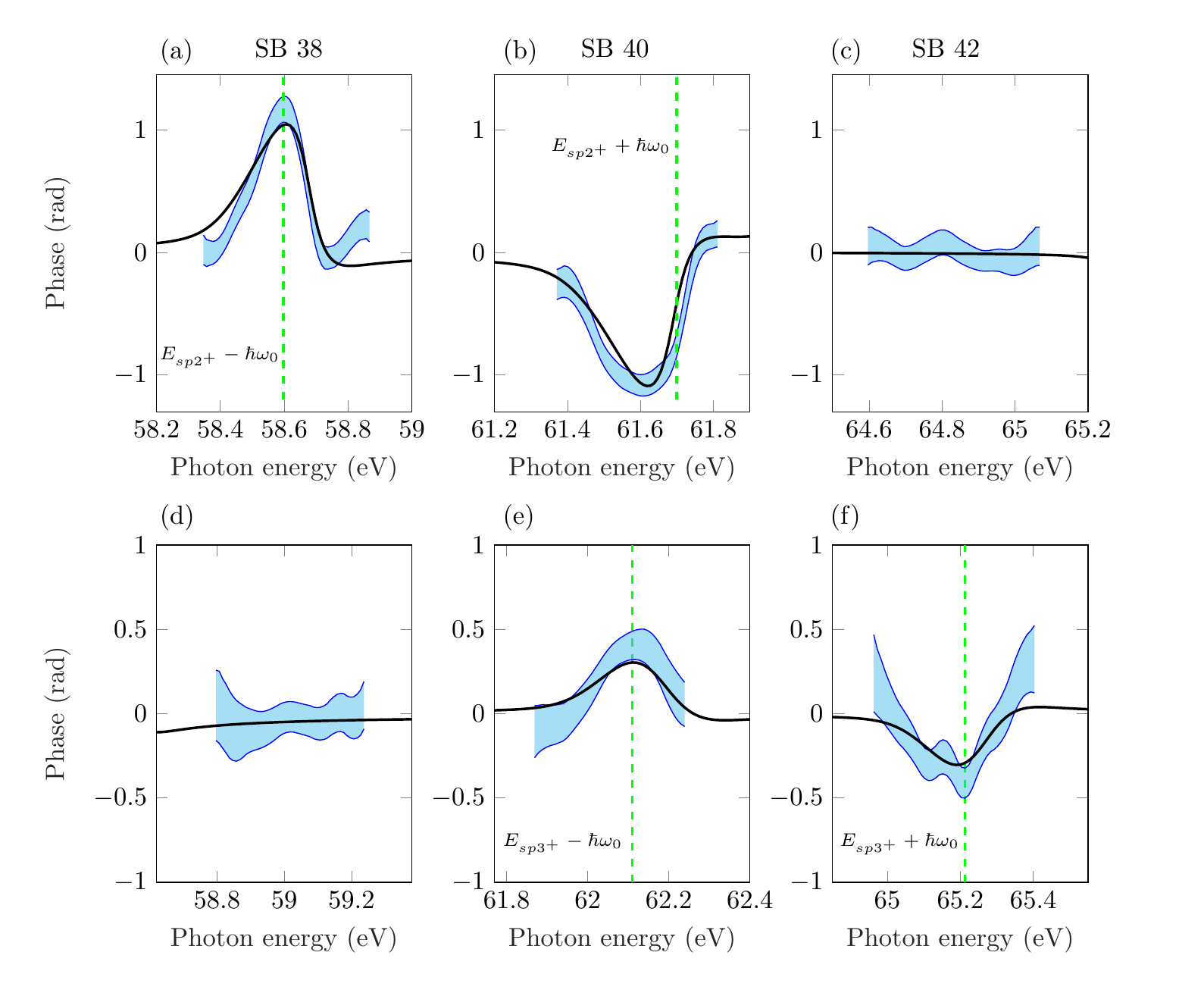}
     \caption{Phases measured (blue) in SB38 (first column), SB40 (second column) and SB42 (third column) in the cases where H39 is resonant with the sp2$^+$ state (first row) and where H41 is resonant with the sp3$^+$ state (second row). The shaded areas correspond to standard deviation around the measured value. A clear phase variation linked to the two resonances is observed on the sidebands originating from the resonant harmonics (SB38 and SB40 for sp2$^+$, SB40 and SB42 for sp3$^+$), while the phase on the other sidebands is flat. The theoretical calculations (black) agree very well with the measured phases.
     }
  \label{fig:phases}
  \end{figure*}

The sp2$^+$ and sp3$^+$ resonances are independently excited by tuning respectively harmonics 39 and 41 to the autoionizing states. When harmonic 39 is resonant with the sp2$^+$ resonance, a clear amplitude modulation of the two-photon wave packet extracted from SB38 is observed [figure \ref{fig:deconv}(a)]. In particular, due to the broad harmonic profile, the amplitude exhibits a double structure which results from the ionization via both resonant and non-resonant continua. The first peak, centered at $E_{\mathrm{f}}=58.6$ eV, and the dip at $E_{\mathrm{f}}=58.7$ eV result from ionization via the sp2$^+$ resonance (green dashed line) and present the typical constructive and destructive interferences characteristic of the Fano profile. The second peak, centered at $E_\mathrm{f}=58.8$ eV, originates from the ionization via a non resonant continuum which is probed by the high energy part of the harmonic.
\textcolor{changes}{When harmonic 41 is resonant with the sp3$^+$ state, the amplitude of $A_{41-1}$ is smoother than that of $A_{39-1}$ in the previous case (figure \ref{fig:sp3}).  The width of the sp3$^+$ resonance (8 meV) is extremely small compared to that of the harmonics (180 meV) and IR pulse (125 meV), and is thus subject to a strong broadening due to the finite pulse effects. This behavior is well reproduced by theory and indicates that, in our experimental conditions, the modification of the \textit{amplitude} of the two-photon wave packet due to the sp3$^+$ resonance cannot be resolved.}

Figure \ref{fig:phases} displays the phases measured for sidebands 38, 40 and 42 when harmonic 39 is resonant with the sp2$^+$ state (upper row) and when harmonic 41 is resonant with the sp3$^+$ state (lower row). As in figure \ref{fig:deconv}(b), we only show the frequency interval such that the phase can be extracted with good accuracy. For both resonances, we can measure a clear phase variation induced by the resonance while the third non resonant sideband, shown for comparison (either SB42 in the first row or SB38 in the second row), exhibits a flat phase, since the two-photon ionization occurs through a smooth continuum. As expected, the phase variations observed for the sidebands where the resonance is one IR photon above or below are similar, except for an opposite sign. For the sp2$^+$resonance, a fast phase variation of 1.2 rad is observed across the resonant part of the sideband. For the sp3$^+$ resonance, despite the smooth amplitude of the resonant wave packet, a phase variation of 0.3 rad is measured, indicating that the EWP is affected by the sp3$^+$ state. However, despite their similar $q$ values, the phase jump measured for the sp3$^+$ is smaller than the one measured for the sp2$^+$ state. This difference  originates from the fact that the width of the sp3$^+$ resonance is almost four times smaller than that of the sp2$^+$ state and is consequently much more sensitive to the finite pulse effects. In general,
phase measurements are more sensitive to the presence of a resonance than amplitude measurements. In the absence of a resonance, the phase is flat, while the amplitude reflects that of the excitation pulse. Phase measurements are thus ``background-free", while amplitude measurements are not.

\begin{table*}[!htbp]
\begin{tabular}{|c|c|c|c|c|c|c|c|}
\hline
 & $\lambda_{\mathrm{IR}} $(nm) & $\sigma_{\mathrm{IR}} $(meV) & $\Delta t_{\mathrm{IR}}$ (fs)& $ I_{\mathrm{IR}}$ (W/cm$^2$)& $\sigma_{\mathrm{XUV}} $(meV) & $I_{\mathrm{XUV}} $ (W/cm$^2$)& $\sigma_{\mathrm{MBES}}$(meV) \\
\hline
This work & 800 & 125 & 30 & 10$^{11}$ &  180& $10^9$ & 89 \\
\hline
\cite{GrusonScience2016}& 1295 & 26 & 70 &$2\times 10^{11}$  & 400 & - & 190 \\
\hline
\end{tabular}
\caption{\textcolor{changes}{Comparison of the experimental parameters between this work and \cite{GrusonScience2016}: central wavelength, bandwidth, pulse duration,  intensity of the IR, bandwidth and intensity of the resonant harmonic, and spectrometer resolution (from left to right).}}
\label{tab:Exp_Parameters}
\end{table*}

Both amplitude and phase measurements are compared to theoretical calculations using the finite pulse model introduced in \cite{GalanPRA2016}. The calculations, which take into account the bandwidths of the harmonics and IR pulses, reproduce well the measured amplitudes [figure \ref{fig:deconv}(a)] and phases (figure \ref{fig:phases}). Furthermore, our measurements are in very good agreement with the ones carried out by Gruson \textit{et al.} \cite{GrusonScience2016}. However, as already mentioned in section \ref{subsec:spectometer}, the limitations to the spectral resolution in the two experiments have different origins.
\textcolor{changes}{Table \ref{tab:Exp_Parameters} summarizes the different experimental parameters in the two experiments. In \cite{GrusonScience2016}, the IR bandwidth was smaller than the resonance width, strongly reducing the influence of the finite-pulse effects so that
the limiting factor was the MBES resolution.}
\textcolor{changes}{In our case, the RABBIT spectrogram is deconvolved from the MBES response but the broad IR bandwidth limits our spectral resolution. Despite the different experimental parameters in the two experiments, the good agreement between the results shows the flexibility of the Rainbow RABBIT technique and its applicability to a wide range of experimental conditions.}

\subsection{Time domain}
\label{sec:time_domain}

The measured spectral amplitude and phase, displayed in figure \ref{fig:deconv}, are now used to reconstruct, using a Fourier transform, the temporal characteristics of the two-photon EWP emitted through the sp2$^+$ resonance. Note that the phase evolution below the 30$\%$-threshold does not affect significantly the reconstruction. Figure \ref{fig:Temporal_profile} shows the temporal intensity (blue solid curve) and phase (blue dotted curve) of the wave packet. The temporal profile shows a large Gaussian-like peak centered at the origin with a duration of 6 fs FWHM reflecting the ionizing XUV pulse. On this time scale, the dominant ionization channel is the direct one. As the autoionizing state decays in the continuum, the contribution of both ionization paths become comparable and strong destructive interferences between the two channels lead to a sharp decrease of the temporal intensity around $t=8$ fs, which is followed by a revival of the EWP. When the intensity profile goes to zero, the temporal phase jumps. The intensity decreases rapidly after 12 fs, much faster than the theoretical lifetime of 17 fs (table \ref{tab:FanoParameters}). This apparently-faster decay of the autoionizing state results from the finite pulse effects and is well reproduced by simulations, indicated by the black line in figure \ref{fig:Temporal_profile} and obtained by Fourier transforming the simulated spectral amplitude [figure \ref{fig:deconv} (a)] and phase [figure \ref{fig:phases}(a)]. It occurs because the short IR pulse probes the decay during a limited amount of time (less than $\simeq$ 15 fs). This observation does not reflect a real modification of the decay rate but is only the result of the lack of spectral resolution. \textcolor{changes}{Note that the good agreement with the theory indicates that the unidentified spectral feature at 58.4 eV does not significantly affect the ionization dynamics.} A similar temporal evolution was obtained in \cite{GrusonScience2016}, with some deviation due to the different experimental conditions \textcolor{changes}{(see table \ref{tab:Exp_Parameters})}.

\begin{figure}[t]
\begin{center}
\includegraphics{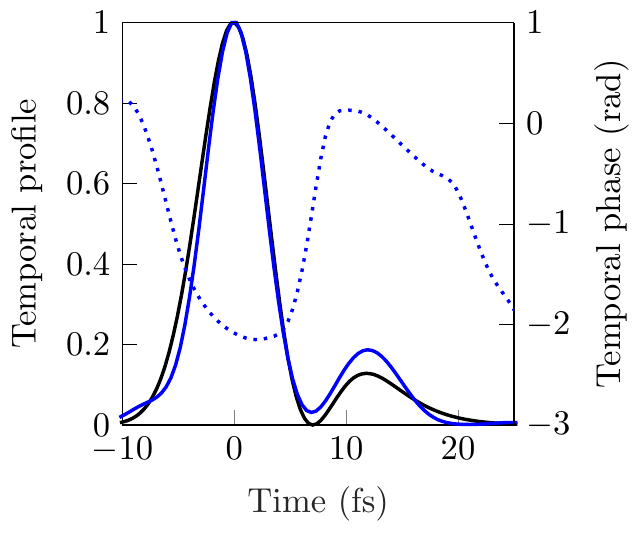}
\end{center}
\caption{Reconstruction of the temporal profile of the resonant two-photon wavepacket. Temporal intensity (blue solid) and phase (blue dotted) of the EWP retrieved from the experiment. The simulated temporal intensity is shown in black line.}
\label{fig:Temporal_profile}
\end{figure}

\begin{figure*}[!htbp]
\begin{center}
\includegraphics{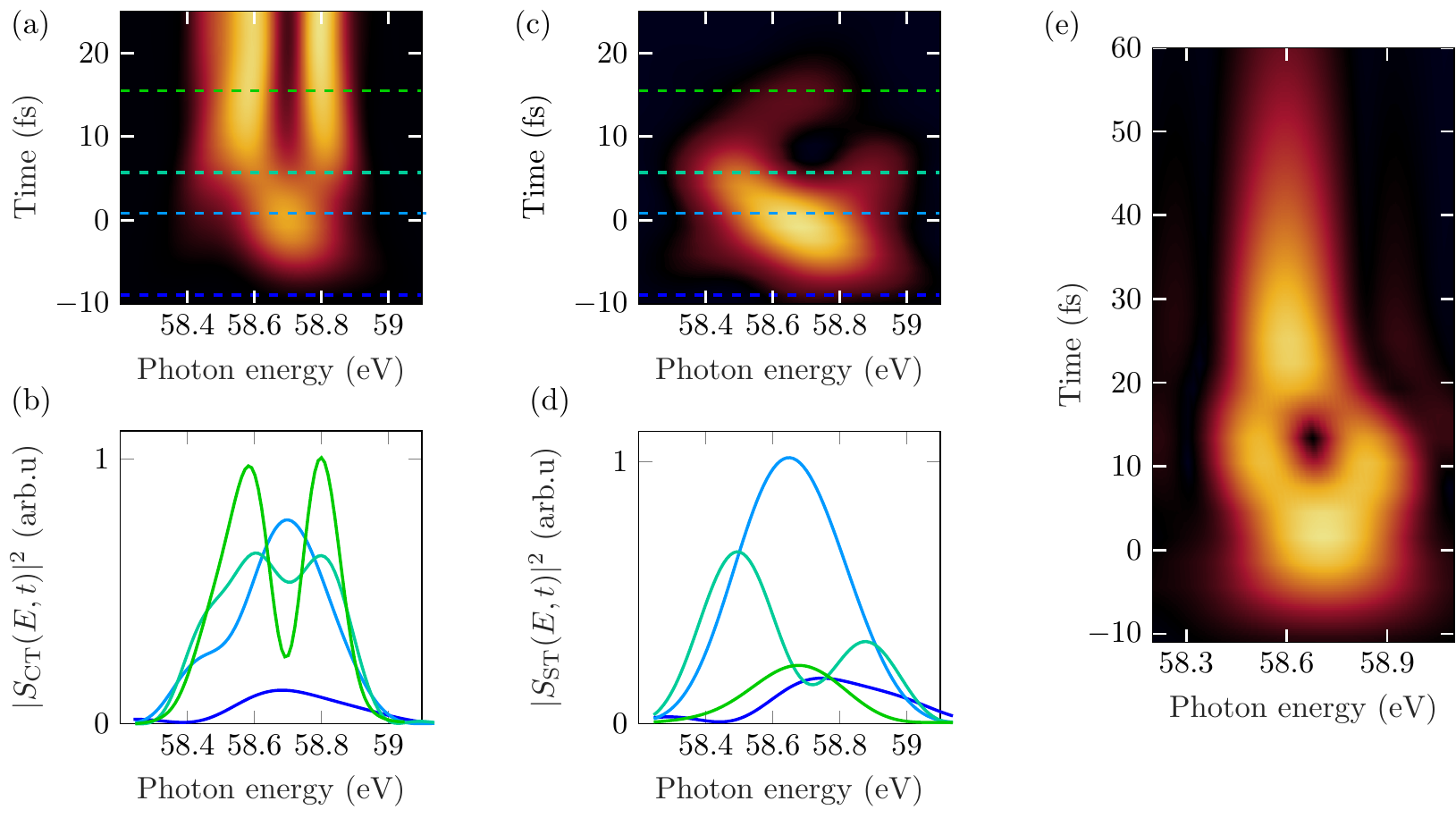}
\end{center}
  \caption{(a) Representation of $|S_\mathrm{C}|^2$ in colors, as a function of $E_f$ and $t$. (b) Lineouts of $|S_\mathrm{C}|^2$ at the times indicated by the dashed lines in (a). (c) Representation of $|S_\mathrm{ST}|^2$ in colors, as a function of $E_f$ and $t$. (d) Lineouts of $|S_\mathrm{ST}|^2$ at the times indicated by the dashed lines in (c). (e) Representation of $|S_\mathrm{ST}|^2$ in colors, as a function of $E_f$ and $t$ using simulated data for a 10 nm broad IR pulse. }
  \label{fig:reconstruction}
\end{figure*}

\section{Time-frequency representation}
\label{sec:time_frequency}

\subsection{Time-limited Fourier transform}
The spectral and temporal domains provide distinct and complementary pictures of the autoionization dynamics. Similarly to what is done in ultrafast optics to characterize optical wave packets, new insights on the ionization process can be gained by representing the evolution of electronic wave packets in the time-frequency space. This can be achieved by using time-frequency representations, e.g. based on inverse Fourier transforms of the complex temporal amplitude of the wave packet with a temporal window. These transforms can be generally written as
\begin{equation}
S(E_\mathrm{f},t)=\int_{-\infty}^{+\infty} \mathrm{d}\uptau\ \tilde{A}_\mathrm{n+1}(\uptau)\alpha(\uptau-t)\exp\left(i\frac{E_\mathrm{f} \uptau}{\hbar} \right),
\label{eq:ft}
\end{equation}
where $\tilde{A}_\mathrm{n+1}(\uptau)$ is the Fourier transform of $A_\mathrm{n+1}(E_\mathrm{f})$ and $\alpha(\uptau-t)$ is the window function used to limit the temporal extent of the Fourier transform. This function can be a Heaviside function [$\alpha(t)=\Theta(-t)$], and we refer to equation \ref{eq:ft} as a cumulative Fourier transform (CFT) \cite{WickenhauserPRL2005} ($S=S_\mathrm{C}$). We also use a Super Gaussian function: $\alpha(t)=\exp[-t^6/(2 \Delta t^6)]$, where $\Delta t$ is the window width (typically 15 fs). In this case, equation \ref{eq:ft} is a short-time Fourier transform (STFT) \cite{CohenIEEE1989} ($S=S_\mathrm{ST}$).

$S_\mathrm{C}(E_\mathrm{f},t)$ represents the spectral amplitude accumulated until time $t$ and its temporal variation shows how the  wave packet builds up in the continuum, as shown in figure \ref{fig:reconstruction}(a) and (b). $S_\mathrm{ST}(E_\mathrm{f},t)$ represents the spectral amplitude emitted within the time window and shows the evolution of the instantaneous frequencies emitted in the continuum [figure \ref{fig:reconstruction}(c) and (d)]. Both representations indicate that during the first 5 fs, a smooth gaussian-like EWP emerges in the continuum. The shape of the wave packet reflects that of the ionizing pulse, revealing that the direct ionization is dominant. Passed this time, the direct and resonant paths start interfering giving rise to destructive interferences at the center of the wave packet (around $E_\mathrm{f}=58.7$ eV) and constructive interferences on both sides. After 8 fs, the two representations start to differ. The STFT shows that the interferences disappear and a weak, spectrally narrow decay is observed around 58.6 eV. The XUV pulse has then passed the interaction region and the atoms cannot be directly ionized. However, the sp2$^+$ state can still decay in the continuum thus giving rise to this weak decay. In contrast, the CFT barely changes after 8 fs because of the small contribution from the decay to the accumulated spectral amplitude.
Finally, figure \ref{fig:reconstruction}(e) shows a STFT obtained from simulations carried out with a long IR pulse. In this case, a decay corresponding to a 17 fs lifetime can be observed.

\subsection{Wigner representation}
The Wigner distribution (WD) is an alternative way of representing the time-frequency structure of the wave packet \cite{BourassinNatComm2015,CohenIEEE1989}. Contrary to the STFT, the WD does not require gating the Fourier transform with an arbitrarily chosen window function. The WD can be defined both in the time and frequency domains as
\begin{eqnarray}
&W(E_\mathrm{f},t)=\int_{- \infty}^{+\infty}\!\!\!\!\!\!\!\! \mathrm{d}\uptau\  \tilde{A}_\mathrm{n+1}\left(t+\frac{\uptau}{2}\right)\tilde{A}_\mathrm{n+1}^*\left(t-\frac{\uptau}{2}\right)e^{iE_\mathrm{f}\uptau/\hbar}\nonumber \\
&=\frac{1}{2\pi}\! \int_{-\infty}^{+\infty}\!\!\!\!\!\!\!\!\!\! \mathrm{d}\varepsilon\ A_\mathrm{n+1}\!\left(E_\mathrm{f}+\frac{\varepsilon}{2}\right)\!A_\mathrm{n+1}^*\!\left(E_\mathrm{f}-\frac{\varepsilon}{2}\right)e^{-i\varepsilon t/\hbar}
\end{eqnarray}
and can be seen as the Fourier transform of the auto-correlation function of the EWP. Additionally, one of the properties of this distribution is that the projections along the time (respectively frequency) axes (referred to as \textit{marginals} in the literature) generates the spectral (respectively temporal) intensity of the wave packet: $\int W(E_\mathrm{f},t) \mathrm{d}t=|A_\mathrm{n+1}(E_\mathrm{f})|^2$ and $2\pi\int W(E_\mathrm{f},t) \mathrm{d}E_\mathrm{f}=|\tilde{A}_\mathrm{n+1}(t)|^2$. Finally, an interesting feature of this representation is that it is not a positive distribution. In the WD of coherent multicomponent signals, the different components interfere with each other and the distribution can take negative values.

\begin{figure}[t]
 \includegraphics{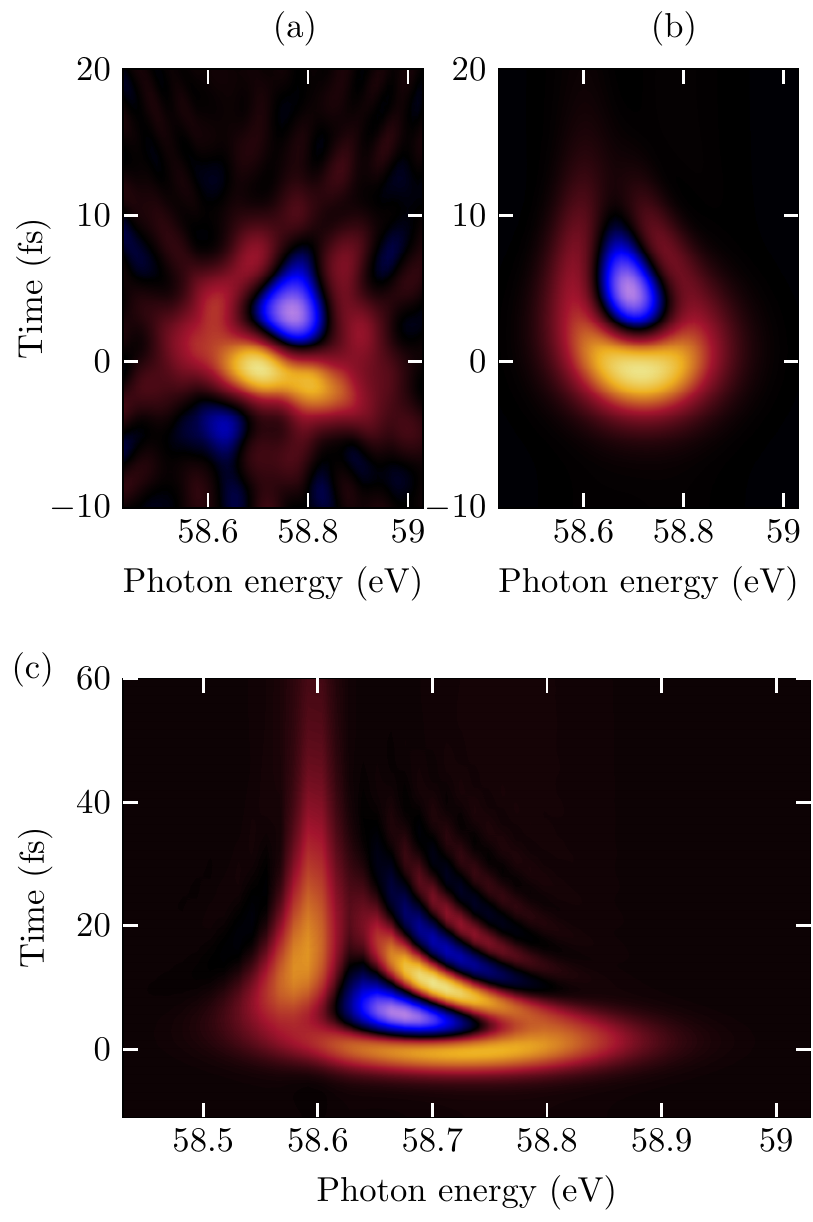}
  \caption{Wigner distribution. (a) Experimental WD, (b) simulated WD using the finite pulse model with experimental parameters for the XUV and IR pulses, (c) simulated WD using the finite pulse model with experimental parameters for the XUV and a 10nm broad IR pulse.}
  \label{fig:Wigner}
\end{figure}

Figure \ref{fig:Wigner}(a) shows the experimental Wigner distribution (WD) of the two-photon EWP emitted through the sp2$^+$ resonance. The spectrally large peak centered at $t=0$ fs represents, like for the STFT, the direct ionization path. The temporally long and spectrally narrow feature centered at $E_\mathrm{f}=58.6$ eV describes the decay of the sp2$^+$ state. Because these two processes have such distinct spectral-temporal representations, it is very easy to disentangle the direct ionization to the continuum states from the autoionization through the sp2$^+$ state. The negative peak and the shoulder between $E_\mathrm{f}=58.7$ eV and $E_\mathrm{f}=58.8$ eV represent the interferences between the two ionization paths.  These results agree very well with the theoretical calculations as shown in figure \ref{fig:Wigner}(b).
These interference effects provide information on the correlation between the direct and resonant ionization amplitudes.
In our experimental conditions, the IR pulses were too short to allow a complete visualization of these correlation effects. In figure \ref{fig:Wigner}(c) we show the simulation of the WD that would be obtained using the same XUV pulses but spectrally-narrower IR pulses of 10 nm bandwidth corresponding to a pulse duration of roughly 100 fs. Very clear oscillations appear between 58.6 and 58.9 eV compared to the simulation in the experimental conditions. These oscillations are characterized by a frequency that increases linearly with the detuning and an amplitude that is damped as a function of time.

\subsection{Analytical Wigner distribution}

In this section we derive analytically the expression of the Wigner distribution for the complex Fano amplitude (equation \ref{eq:resonant_factor}). We first take the Fourier transform as in \cite{GrusonScience2016,ZhaoPRA2005}:
\begin{equation}
\tilde{R}(t)=\delta(t)-i\frac{\Gamma}{2\hbar}(q-i) \: \mathrm{e}^{-\left(iE_\Phi/\hbar+\Gamma/2\hbar\right)t} \: \Theta(t)
\end{equation}
The Wigner distribution can be written as the sum of three terms, $W(E,t)= W_D +W_I+ 2\mathrm{Re} \left(W_{ID}\right)$, which are defined as:
\begin{eqnarray}
W_D(E,t) = \int \delta(t+\frac{\tau}{2}) \delta(t-\frac{\tau}{2}) \: \mathrm{e}^{iE\tau/\hbar} \: \mathrm{d}\tau
\end{eqnarray}
\begin{eqnarray}
&W_I(E,t) = \frac{\Gamma^2}{4\hbar^2} \int (q-i)\mathrm{e}^{-\left(iE_\Phi/\hbar+\Gamma/2\hbar\right)\left(t+\frac{\tau}{2}\right)}\Theta\left(t+\frac{\tau}{2}\right)\nonumber \\
&\times (q+i)\mathrm{e}^{\left(iE_\Phi/\hbar-\Gamma/2\hbar\right)\left(t-\frac{\tau}{2}\right)}\Theta\left(t-\frac{\tau}{2}\right) \mathrm{e}^{iE\tau/\hbar} \: \mathrm{d}\tau
\end{eqnarray}
\begin{eqnarray}
&W_{ID}(E,t)=\int \mathrm{d}\tau \: \delta(t+\frac{\tau}{2})\ i\frac{\Gamma}{2\hbar}(q+i) \nonumber \\ & \times \exp\left[\left(\frac{iE_\Phi}{\hbar}-\frac{\Gamma}{2\hbar}\right)\left(t-\frac{\tau}{2}\right)\right]\Theta\left(t-\frac{\tau}{2}\right) \mathrm{e}^{iE\tau/\hbar}
\end{eqnarray}

The calculation of $W_D$ is straightforward and gives $W_D=\delta(t)$. For $W_I$ we get:
\begin{equation}
W_I=\frac{\Gamma^2}{2\hbar}(q^2+1)\exp\left(-\frac{\Gamma t}{\hbar}\right)\frac{\sin\left(2\frac{E-E_\Phi}{\hbar}t\right)}{E-E_\Phi}
\end{equation}
Finally we get that
\begin{eqnarray}
& 2\mathrm{Re}(W_{ID})=\frac{\Gamma}{\hbar}e^{-\Gamma t /\hbar} \: \Theta(t) \nonumber \\
& \times \left[q\sin\left(\frac{2(E-E_\Phi)t}{\hbar}\right)-\cos\left(\frac{2(E-E_\Phi)t}{\hbar}\right)\right]
\end{eqnarray}
The first term, $W_D$, corresponds to the direct transition to the continuum and, once convolved with the spectrum of the ionizing pulse, results in a large feature observed at $t$ = 0 fs. The second term $W_I$, describes the decay of the autoionizing state in the continuum, at the energy $E_\Phi$. Finally the last term results from the interference of both contributions and leads to oscillations with an hyperbolic shape observed in figure \ref{fig:Wigner}(c).

\section{Conclusion}
\label{sec:conclusion}

In summary, we have presented calculations and measurements of the amplitude and phase of EWPs emitted through the sp2$^+$ and sp3$^+$ Fano resonances in helium using the Rainbow RABBIT technique. We discussed aspects that may affect these spectrally-resolved measurements, in particular the spectrometer resolution and the finite pulse effects. The retrieved amplitude and phase were then used to fully characterize the wave packets in the time-frequency space which allowed us to disentangle the dynamics associated with the different ionization channels. The sensitivity of the technique can be improved by using harmonics with a broad spectral width providing a locally smooth amplitude variation and allowing the measurement of fast changes in the ionization cross section. In addition, the combination of long dressing pulses to minimize finite-pulse effects with photoelectron spectrometers with high resolving power (combined with deconvolution techniques) can greatly increase the spectral resolution of the RABBIT technique. Time-frequency representations offer a powerful tool to characterize EWP emitted close to resonances where strong electron correlations lead to significant time-frequency couplings. In particular, we have shown that the WD can potentially provide unique insights into electron correlation during autoionization. This could be extended to the study of photoionization in more complex systems such as molecules of biological interest and solids. Another potential application of the WD is the complete tomographic reconstruction of partially coherent ultra-short pulses in the XUV and X-ray range \cite{BourassinNatComm2015}.

\section*{Acknowledgments}
This research was supported by the European Research Council (Advanced grant PALP), the Swedish Research Council, the Knut and Alice Wallenberg Foundation, Sweden, the Laserlab-Europe EU-H2020 654148, the EU H2020-MSCA-ITN-MEDEA-641789, the ANR-15-CE30-0001-CIMBAAD, ANR-11-EQPX0005-ATTOLAB and ANR-10-LABX-0039-PALM.

\section*{References}
\bibliographystyle{unsrt}
\bibliography{Bibliography.bib}

\end{document}